\documentclass[11pt]{article}
\usepackage{hyperref}
\usepackage{amsmath}
\usepackage{amssymb}
\usepackage{bm}
\usepackage{latexsym}
\usepackage{graphicx}
\usepackage{euscript} 
\usepackage{mathrsfs}
\begin{document}
\title{Evolution of the spectral index after inflation}
\author{Ali A. Asgari \\Amir H. Abbassi*\\
\\\vspace{6pt} Department of Physics, School of Sciences,\\ Tarbiat Modares University, P.O.Box 14155-4838, Tehran, Iran\\}
\maketitle
%%%The abstract (in this file, and that submitted as text to arXiv) should include the exact phrase
\begin{abstract}
In this article we investigate the time evolution of the adiabatic (curvature) and isocurvature (entropy) spectral indices after end of inflation for all cosmological scales and two different initial conditions. For this purpose, we first extract an explicit equation for the time evolutin of the comoving curvature perturbation (which may be known as the generalized Mukhanov-Sasaki equation). It shall be manifested that the evolution of adiabatic spectral index severely depends on the initial conditions and just for the super-Hubble scales and adiabatic initial conditions is constant as be expected. Moreover, it shall be clear that the adiabatic spectral index after recombination approach to a constant value for the isocurvature perturbations. Finally, we re-investigate the Sachs-Wolfe effect and show that the fudge factor $ \frac{1}{3} $ in the adiabatic ordinary Sachs-Wolfe formula must be replaced by 0.4.
\end{abstract}
%%%The main text
\section{Introduction}
\label{sec:intro}

Inflation was produced first for resolving three classical cosmological problems: the horizon, flatness and monopole problems \cite{1}. It also explains the origin of the CMB anisotropy and structure formation \cite{2,3}, indeed, during inflation the quantum vacuum fluctuations of the scalar field(s) on the scales less than the Hubble horizon are magnified into the classical perturbations in the scalar field(s) on scales larger than the Hubble horizon \cite{4,5}. It can be shown that these perturbations have the nearly scale-invariant spectrum and can explain the origin of the inhomogeneities in the recent universe such as large structures and CMB anisotropies as well \cite{2,6}. These classical perturbations can be described by some perturbative cosmic potentials which are related to the FLRW metric perturbations and maybe cosmic fluid perturbations. One of these perturbative cosmic potentials is the \textit{comoving curvature perturbation} $ \mathcal{R} $ which is significant in cosmology due to the following reasons
\begin{itemize}
\item It is conserved for the adiabatic perturbations when the scales of the perturbations are extremely longer than the Hubble horizon \cite{6}
\item Gauge-invariance and resemblance to the physical observables \cite{7}
\item Sasaki-Stewart $\delta N$-formula which express the perturbation of e-folding number in
terms of $ \mathcal{R} $ \cite{8}
\item The scalar primordial power spectrum usually refers to the power spectrum of $ \mathcal{R} $ which characterizes the adiabatic mode \cite{5,9,10}
\end{itemize}
Furthermore, it can be shown that the linear perturbation of the scale factor and signature of the spatial curvature of the universe in the comoving gauge merely depends on $ \mathcal{R} $ \cite{11,12,13}
\begin{eqnarray*}
&&\delta a\left(t,\textbf{x} \right)= a\left(t \right) \mathcal{R}\left(t,\textbf{x} \right),\\ \\
&&\delta K\left(t,\textbf{x} \right)=-\frac{2}{3} \nabla ^2\mathcal{R}\left(t,\textbf{x} \right).
\end{eqnarray*}
Finally, $ \mathcal{R} $ can be used to connect observed cosmological perturbations in the adiabatic mode with quantum fluctuations at very early times \cite{6}. The dynamic of $ \mathcal{R} $ during the inflation is described by the well-known \textit{Mukhanov-Sasaki equation} \cite{14,15}. This equation yields the power spectrum and spectral index of $ \mathcal{R} $ at the inflation era. In this paper we generalize the Mukhanov-Sasaki equation to be included all the history of the universe and then by invoking a simple model, show that how $ \mathcal{R} $ evolves after inflation. We also discuss about the spectral index evolution after the inflation.\\
The outline of this paper is as follows. In Section 2, we present an explicit equation for the  time evolution of $ \mathcal{R} $ which can be used for all history eras of the universe and then investigate its solutions for some very simple cases. In Section 3, we investigate a universe which contains mixture of radiation and matter and then show that the $ \mathcal{R} $-evolution equation can be solved after coupling to the\textit{Kodama-Sasaki equation} \cite{16,17}. We consider two adiabatic and isocurvature initial conditions and present the numerical solutions. We also supply an analytic method which helps us to approximate solutions. Moreover, we present the numerical results of the curvature spectral index and also entropy spectral index evolution in the post-inflationary universe. In Section 4, we re-investigate the Sachs-Wolfe effect and point out that $ \frac{1}{3} $ factor in the
Sachs-Wolfe formula must be enhanced as will be said. We present our conclusion in Section 5.
%%%%%%%%%%%%%%%%%%%%%%%%%%%%%%%%%%%
\section{A general equation for evolution of $ \mathcal{R}_q $}
\label{sec:1}

We may consider the metric of the universe as \cite{6,18,19}
\begin{equation}\label{a1}
ds^2=a^2\left\lbrace − \left( 1 + E\right)d\tau ^2+ 2\partial_i F d\tau dx^i+ \left[ \left( 1 + A\right) \delta_{ij} + \partial_i\partial_j B\right] dx^idx^j\right\rbrace ,
\end{equation}
which is the FLRW metric with $ K=0 $  in the comoving quasi-Cartesian coordinates accompanied by the most general scalar linear perturbations. Similarly, the energy-momentum tensor of the cosmic fluid can be written as
\begin{eqnarray}
&&T_{00}= a^2\left[ \bar \rho\left( 1 + E\right)  + \delta\rho\right] ,\\
&&T_{i0} = a^2\left[\bar p \partial_i F-\left(\frac{\bar \rho+\bar p}{a} \right)\partial _i\left(\delta u \right) \right] ,\\
&&T_{ij} = a^2\left[\bar p\left(1+A \right)\delta _{ij}+\delta p \delta _{ij}+\bar p\partial _i\partial _j B +\partial  _i\partial _j \Pi ^S \right] , \label{a2}
\end{eqnarray}
where $ \delta u $ and $ \Pi^S $ are respectively the scalar velocity potential and scalar anisotropic inertia of the cosmic fluid. $ \Pi^S $ represents departures of the cosmic fluid from perfectness. Furthermore, $ \rho=\bar\rho+\delta\rho $ and $ p=\bar p+\delta p $ are the energy density and pressure of the cosmic fluid respectively. Notice that bar over every quantity stands for its unperturbed value. It can be shown that except $ \Pi^S $ all of the perturbative scalars in equations (\ref{a1}) to (\ref{a2}) are not gauge-invariant \cite{6}, so we may invoke combinational gauge-invariant scalars like Bardeen potentials \cite{7}
\begin{eqnarray}
&&\Psi = −\frac{A}{2}-\mathcal{H}\sigma ,\\
&&\Phi =\frac{E}{2}+\mathcal{H}\sigma +\sigma' .
\end{eqnarray}
Here the prime symbol stands for the derivative respect to the $ \tau $ and $ \mathcal{H}=Ha $ is the
comoving Hubble parameter. Furthermore, $ \sigma=F-\frac{1}{2}B' $ is the shear potential of the cosmic fluid. Some other gauge-invariant scalars are
\begin{eqnarray}
&&\mathcal{R} =\frac{A}{2}+H\delta u,\\
&&\zeta =\frac{A}{2}-\mathcal{H}\frac{\delta\rho}{\bar \rho'},\\
&&\Delta =a\frac{\delta\rho}{\bar \rho'}+\delta u,\\
&&\Gamma=\delta p-{c_s}^2\delta\rho ,
\end{eqnarray}
where $ c_s^2 $ is the adiabatic sound speed in the cosmic fluid. $ \mathcal{R} $ is known as comoving (intrinsic) curvature perturbation. According to the perturbative form of the field equations and also energy-momentum conservation law we can write
\begin{eqnarray}
\mathcal{R}_q&=&\frac{-2\mathcal{H}^2+\mathcal{H}'}{\mathcal{H}^2-\mathcal{H}'}\Psi _q-\frac{\mathcal{H}}{\mathcal{H}^2-\mathcal{H}'}\Psi'_q +\frac{8\pi G\mathcal{H}^2a^2}{\mathcal{H}^2-\mathcal{H}'}\Pi^S_q ,\label{a3}\\
\mathcal{R}'_q&=&\frac{\mathcal{H}{c_s}^2}{\mathcal{H}^2-\mathcal{H}'}q^2\Psi _q-\frac{4\pi G\mathcal{H}a^2}{\mathcal{H}^2-\mathcal{H}'}\Lambda_q ,\label{a4}
\end{eqnarray}
where $ \mathcal{R}_q $ denotes the Fourier transformation of $ \mathcal{R} $ with the comoving wave number $ q $ and $ \Lambda _q=\Gamma_q-q^2\Pi^S_q $. Notice that we have taken $ K=0 $ which is in agreement with the observational data\cite{20}. By combination of equations (\ref{a3}) and (\ref{a4}) after some tedious but straightforward calculation we can derive an explicit equation in terms of $ \mathcal{R}_q $
\begin{equation}\label{a5}
\mathcal{R}'_q=-\frac{4\pi G\mathcal{H}a^2}{\mathcal{H}^2-\mathcal{H}'}\Lambda_q ,\quad {c_s}^2=0 \end{equation}
and
\begin{multline}\label{a6}
\mathcal{R}''_q+2\frac{\EuScript{Z}'}{\EuScript{Z}}\mathcal{R}'_q+{c_s}^2q^2\mathcal{R}_q=\\
\qquad\frac{4\pi Ga^2}{\mathcal{H}^2-\mathcal{H}'}\Bigg[\Bigg( -4\mathcal{H}^2+\frac{\left(\mathcal{H}{c_s}^2 \right)' }{{c_s}^2}\Bigg) \Lambda _q- \mathcal{H}\Lambda' _q-2q^2\mathcal{H}^2{c_s}^2\Pi^S _q\Bigg]  ,  \quad {c_s}^2\neq 0
\end{multline}
where $ \EuScript{Z}=\frac{a}{\mathcal{H}}\sqrt{\frac{|\mathcal{H}^2-\mathcal{H}'|}{{c_s}^2}} $. Equations (\ref{a5}) and (\ref{a6}) together with each other include the most general case of the fluid even the scalar field, so the Mukhanov-Sasaki equation is a special case of this equation. For the pure dust universe equation (\ref{a5}) yields  $ \mathcal{R}_q=const $. It means $ \mathcal{R}$ is conserved if the cosmic fluid is dust regardless of the comoving wave number scale. On the other hand, for the pure radiation case equation (\ref{a6}) reduces to $ \mathcal{R}''_q+\frac{q^2}{3}\mathcal{R}_q=0 $ and consequently, $ \mathcal{R}_q \propto \cos\left(\frac{q}{\sqrt{3}}\tau \right) $. Besides, if we suppose the radiation era starts immediately after the inflation, it may apply the following initial condition \cite{6}
\begin{displaymath}
\tau\longrightarrow 0 \qquad :\qquad  \mathcal{R}_q \longrightarrow Nq^{-2+\frac{n_{s_0}}{2}}\quad\left( N \simeq 10^{-5}\quad \rm{and} \quad n_{s_0}\simeq 0.96\right) 
\end{displaymath}
( $ \tau =0 $ is supposed to be the end of inflation era and start time of the radiation epoch.) Thus 
\begin{equation}
\mathcal{R}_q\left(\tau \right) =Nq^{-2+\frac{n_{s_0}}{2}}\cos\left(\frac{q}{\sqrt{3}}\tau \right) .
\end{equation}
Now lets turn to the inflaton case. Every scalar field can be treated as a perfect fluid. For the homogeneous inflaton field $ \bar \varphi\left( t \right)  $ with potential $ V\left(\bar \varphi \right)  $ we have
\begin{eqnarray}
&&\bar\rho=\frac{1}{2a^2}\bar\varphi'{} ^2+ V \left(  \bar\varphi\right) ,\\
&&\bar p =\frac{1}{2a^2}\bar\varphi'{} ^2- V \left(  \bar\varphi\right) .
\end{eqnarray}
It can be shown that under the linear perturbations of the metric i.e. equation (\ref{a1})
\begin{eqnarray}
&&\delta\rho = -\frac{1}{2a^2}E\bar\varphi'{} ^2+\frac{1}{a^2}\bar\varphi'\delta\varphi' +\frac{\partial V}{\partial \bar\varphi}\delta\varphi ,\\
&&\delta p = -\frac{1}{2a^2}E\bar\varphi'{} ^2+\frac{1}{a^2}\bar\varphi'\delta\varphi' -\frac{\partial V}{\partial \bar\varphi}\delta\varphi.
\end{eqnarray}
Now lets confine ourselves to the comoving gauge which indicates $ \delta\varphi=0 $, thus $ \delta\rho=\delta p $, consequently $ {c_s}^2=1 $ and $ \Gamma_q=\Lambda_q=0 $. Note that anisotropic inertia $ \Pi^S_q $ for the scalar fields vanishes. So equation (\ref{a6}) reduces to
\begin{equation}\label{a7}
\mathcal{R}''_q+2\frac{\EuScript{Z}'}{\EuScript{Z}}\mathcal{R}'_q+q^2\mathcal{R}_q=0.
\end{equation}
On the other hand, according to the Friedmann equation $ \mathcal{H}^2-\mathcal{H}'=4\pi G\bar\varphi ^2 $ Thus $ \EuScript{Z}=\frac{a\varphi'}{\mathcal{H}} $. Now by introducing the Sasaki-Mukhanov variable as $ v_q=\EuScript{Z}\mathcal{R}_q $, equation(\ref{a7}) reduces to
\begin{equation}
v''_q+\left(q^2-\frac{\EuScript{Z}''}{\EuScript{Z}} \right)v_q=0 ,
\end{equation}
which is the famous Mukhanov-Sasaki equation.
%%%%%%%%%%%%%%%%%%%%%%%%%%%%%%%%%%
\section{Evolution of $ \mathcal{R}_q $ in a simplified universe}
\label{sec:2}

\begin{figure}[!htb]
\centering
\includegraphics[width=.6\textwidth]{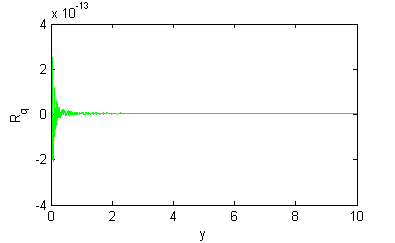}
\hfill
\includegraphics[width=.6\textwidth]{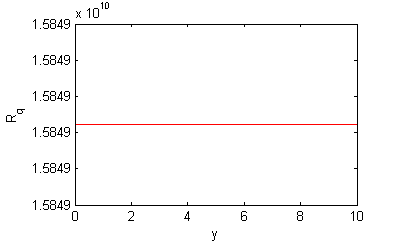}
\caption{Evolution of the comoving curvature perturbation $ \mathcal{R}_q $ in a universe constructed from dust and radiation for the comoving wave number $ q=10^5 $ i.e. sub-horizon scales (up) and $ q=10^{-10} $ namely severe super-horizon modes (down) providing the adiabatic initial condition. We suppose $ n_{s_0}=0.96 $ and $ N $, the amplitude of $ \mathcal{R}_q $ at the end of inflation is roughly $ 10^{-5} $. It is clear that $ \mathcal{R}_q $ for the super-horizon scales is conserved. Notice that both $q$ and $ \mathcal{R}_q $ are dimensionless.}\label{fig1}
\end{figure}

\begin{figure}[!htb]
\centering
\includegraphics[width=.6\textwidth]{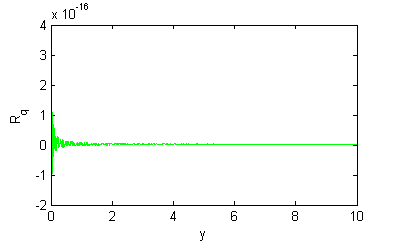}
\hfill
\includegraphics[width=.6\textwidth]{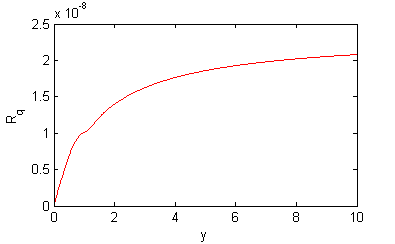}
\caption{The same as Figure \ref{fig1}, exept the initial condition has changed to the isocurvature ones. We supposed the amplitude of $ \mathcal{S}_q $ at the end of inflation is about $ 10^{-5} $ and $ n_{iso_0}=4.43 $.}\label{fig2}
\end{figure}

In this section we consider the case where the cosmic fluid has been constructed from two perfect fluids: matter and radiation which dont interact with each other, i.e., there is no energy or momentum transfer between them. This model was introduced first by Peebles and Yu \cite{21} and is used by Seljak \cite{22} in order to approximate CMB anisotropy. Compared to the real universe this is a simplification since, contrary to the CDM, the baryonic matte does interact with photons. Indeed, the radiation components of the real universe i.e. neutrinos and photons behave like a perfect fluid only until their decoupling \cite{17}.
We have now two fluid components, so $ \rho=\rho_R+\rho_M $ where $ \bar\rho_M\propto\frac{1}{a^3} $ and $ \bar\rho_R\propto\frac{1}{a^4} $. Lets define the normalized scale factor as
\begin{equation}
y=\frac{a}{a_{eq}}=\frac{\bar\rho_M}{\bar\rho_R},
\end{equation}
where $ a_{eq} $ is the scale factor in the time of matter-radiation equality. It is clear that
\begin{eqnarray*}
&&\bar\rho_M=\frac{y}{y+1}\bar\rho , \\
 &&\bar\rho_R=\frac{1}{y+1}\bar\rho .
\end{eqnarray*}
Consequently,
\begin{equation}\label{a8}
\omega =\frac{1}{3\left( y+1\right) }, \hspace{35pt} {c_s}^2=\frac{4}{3\left( 3y+4\right) },
\end{equation}
Also
\begin{equation}\label{a9}
\mathcal{H} =\frac{y'}{y}=\frac{\mathcal{H}_{eq}}{\sqrt{2} }\frac{\sqrt{y+1}}{y},
\end{equation}
where $ \mathcal{H}_{eq} $ is the comoving Hubble parameter in the time of matter-radiation equality.
It is not hard to show that
\begin{equation}
\Gamma =-\bar\rho_M{c_s}^2\mathcal{S} ,
\end{equation}
where $ \mathcal{S}=\delta_M-\delta_R=\frac{\delta\rho_M}{\bar\rho_M}-\frac{3}{4}\frac{\delta\rho_R}{\bar\rho_R} $ is the entropy perturbation between matter and radiation. Thus
\begin{equation}\label{b1}
\Lambda _q =\Gamma _q =-\bar\rho _{M_{eq}}\frac{4\mathcal{S}_q}{3y^3\left( 3y + 4\right) }=-\frac{\mathcal{H}_{eq}^2}{4\pi Ga_{eq}^2}\frac{\mathcal{S}_q}{y^3\left( 3y + 4\right)} ,
\end{equation}
By substituting equations (\ref{a8}) , (\ref{a9}) and (\ref{b1}) in equation (\ref{a6}) we find
\begin{equation}\label{b2}
\left( y+1\right) \mathcal{R}_q^{\ast\ast}+\frac{21y^2+36y+16}{2y\left( 3y+4\right) }\mathcal{R}_q^\ast+\frac{8}{3\left( 3y+4\right) }\left( \frac{q}{\mathcal{H}_{eq}}\right) ^2\mathcal{R}_q=\frac{2}{y\left( 3y+4\right) }\mathcal{S}_q+\frac{4\left( y+1\right) }{\left( 3y+4\right)^2 }\mathcal{S}_q^\ast ,
\end{equation}
where "$ \ast $" stands for the partial derivative respect to $ y $.\\
The adiabatic solution of equation (\ref{b2}) for the super-Hubble scales may be found by putting $ \mathcal{S}_q=0 $ and neglecting the term contains $ \frac{q}{\mathcal{H}_{eq}} $ i.e.
\begin{equation}\label{b3}
\mathcal{R}_q^o=C_1\left( q\right) \left( \ln \frac{\sqrt{y+1}+1}{\sqrt{y}}-\frac{\sqrt{y+1}}{y}\frac{3y+2}{3y+4}\right) +C_2\left( q\right) . 
\end{equation}
The first term of equation (\ref{b3}) is decaying through the time and has no any significance in the late times, so we conclude the conservation of $ \mathcal{R} $ as it is be expected \cite{6}. In order to solve equation (\ref{b2}) in general, we ought to determine $\mathcal{S}_q $.  $\mathcal{S}_q $ for a mixture of matter and radiation may be obtained from Kodama-Sasaki equation\cite{16}
\begin{equation}\label{b4}
\frac{1}{\mathcal{H}^2}\mathcal{S}''_q+3\frac{{c_s}^2}{\mathcal{H}}\mathcal{S}'_q=-\frac{q^2}{\mathcal{H}^2}\left[ \tilde{\Delta}_q-\frac{1}{3}\left(3{c_s}^2-1 \right) \mathcal{S}_q\right], 
\end{equation}
where $ \tilde{\Delta}=H\Delta $. By re-writing equation (\ref{b4}) in terms of $ y $ we find
\begin{eqnarray}\label{b5}
y\mathcal{S}_q^{\ast\ast}+\left( 1+\frac{4}{3y + 4}-\frac{y+2}{2\left( y+1\right) }\right) \mathcal{S}_q^\ast=-\frac{q^2}{\mathcal{H}_{eq}^2}\frac{2y}{y+1}\left( \tilde{\Delta}_q+\frac{y}{3y+4}\mathcal{S}_q\right). 
\end{eqnarray}
On the other hand, we have
\begin{equation}\label{b6}
\mathcal{R}'=-\frac{\mathcal{H}{c_s}^2}{\mathcal{H}^2-\mathcal{H}'}\nabla^2\Psi -\frac{4\pi G\mathcal{H}a^2}{\mathcal{H}^2-\mathcal{H}'}\Lambda .
\end{equation}
Furthermore, from Poisson's equation we can write
\begin{equation}\label{b7}
\nabla^2\Psi=-12\pi G\left( \bar\rho+\bar p\right) a^2\tilde{\Delta}.
\end{equation}
Combination of equations (\ref{b6}) and (\ref{b7}) yields
\begin{equation}\label{b8}
\mathcal{R}_q^\ast =\frac{4}{y\left( 3y+4\right) }\left( \tilde{\Delta}_q+\frac{y}{3y+4}\mathcal{S}_q\right) .
\end{equation}
Substituting equation (\ref{b8}) in equation (\ref{b5}) yields
\begin{equation}\label{b9}
\left( y+1\right) \mathcal{S}_q^{\ast\ast}+\frac{3y^2+12y+8}{2y\left( 3y+4\right) }\mathcal{S}_q^\ast=-\frac{1}{2}\left( \frac{a}{\mathcal{H}_{eq}}\right) ^2y\left( 3y+4\right) \mathcal{R}_q^\ast .
\end{equation}
Equations (\ref{b2}) and (\ref{b9}) are coupled and must be solved simultaneously

\begin{eqnarray*}
\left\{\begin{aligned}
&\left( y+1\right) \mathcal{S}_q^{\ast\ast}+\frac{3y^2+12y+8}{2y\left( 3y+4\right) }\mathcal{S}_q^\ast =-\frac{1}{2}\epsilon y\left( 3y+4\right) \mathcal{R}_q^\ast ,\\ \\
&\left( y+1\right) \mathcal{R}_q^{\ast\ast}+\frac{21y^2+36y+16}{2y\left( 3y+4\right) }\mathcal{R}_q^\ast+\frac{8\epsilon}{3\left( 3y+4\right) }\mathcal{R}_q=
\frac{2}{y\left( 3y+4\right) }\mathcal{S}_q+\frac{4\left( y+1\right) }{\left( 3y+4\right)^2 }\mathcal{S}_q^\ast,
\end{aligned}\right.
\end{eqnarray*}
where $ \epsilon=\left( \frac{a}{\mathcal{H}_{eq}}\right)^2 $. At early times, all cosmologically intersecting scales are outside the Hubble horizon, so we may find the behavior of the solutions in the early stage by setting $ \epsilon=0 $. In this case we will have two independent solutions
\begin{itemize}
\item {\bf Solution 1}
\begin{eqnarray*}
\left\{\begin{aligned}
&\mathcal{S}_q=0,\\ \\
&\mathcal{R}_q=const.
\end{aligned}\right.
\end{eqnarray*}
\item {\bf Solution 2}
\begin{eqnarray*}
\left\{\begin{aligned}
&\mathcal{S}_q=const,\\ \\
&\mathcal{R}_q=\frac{y}{3y+4}\mathcal{S}_q=\frac{1}{3}\left( 1-3{c_s}^2\right) \mathcal{S}_q.
\end{aligned}\right.
\end{eqnarray*}
\end{itemize}
Solutions 1 and 2 are called adiabatic and isocurvature initial conditions respectively. Note that they are solutions of the equations (\ref{b2}) and (\ref{b9}) only in the early times when the scales of the perturbations have been extremely longer than the Hubble horizon. The "initial condition" expression refers to this point.\\
From equation(\ref{b2}) it is clear that the entropy perturbation is a source for the curvature perturbation, so if $ \mathcal{S}\neq0 $, $ \mathcal{R} $ is never conserved. In order to solve equations (\ref{b2}) and (\ref{b9}) analytically, we may expand $\mathcal{S}_q $ and $ \mathcal{R}_q $ in terms of $  \epsilon $ by the Frobenius method\cite{23}
\begin{eqnarray}
&&\mathcal{S}_q\left( y\right) =\epsilon^\alpha \sum_{n=0}^\infty {\epsilon^n\mathcal{S}_n\left( y\right)}\label{c1},\\
&&\mathcal{R}_q\left( y\right)=\epsilon^\beta \sum_{n=0}^\infty {\epsilon^n\mathcal{R}_n\left( y\right),}\label{c2}
\end{eqnarray}
Where $  \alpha $ and $ \beta $ , in general, are two arbitrary complex numbers. After substituting equations (\ref{c1}) and (\ref{c2}) in equations (\ref{b2}) and (\ref{b9}) and also setting $ \alpha=\beta $ we have
\begin{eqnarray}\label{c3}
\left\{\begin{aligned}
&\left( y+1\right) \mathcal{S}_0^{\ast\ast}+\frac{3y^2+12y+8}{2y\left( 3y+4\right) }\mathcal{S}_0^\ast=0.\\ \\
&\left(y+1\right) \mathcal{R}_0^{\ast\ast}+\frac{21y^2+36y+16}{2y\left( 3y+4\right) }\mathcal{R}_0^\ast=\frac{2}{y\left( 3y+4\right) }\mathcal{S}_0+\frac{4\left( y+1\right) }{\left( 3y+4\right)^2 }\mathcal{S}_0^\ast.
\end{aligned}\right.
\end{eqnarray}
And also for $ n\geq1 $ the we find two recursive equations as follows
\begin{eqnarray}\label{c4}
\left\{\begin{aligned}
&\left( y+1\right) \mathcal{S}_n^{\ast\ast}+\frac{3y^2+12y+8}{2y\left( 3y+4\right) }\mathcal{S}_n^\ast=-\frac{1}{2}y\left( 3y+4\right) \mathcal{R}_{n-1}^\ast .\\  \\
&\left( y+1\right) \mathcal{R}_n^{\ast\ast}+\frac{21y^2+36y+16}{2y\left( 3y+4\right) }\mathcal{R}_n^\ast=-\frac{8}{3\left( 3y+4\right)}\mathcal{R}_{n-1}+\frac{2}{y\left( 3y+4\right) }\mathcal{S}_n+\frac{4\left( y+1\right) }{\left( 3y+4\right)^2 }\mathcal{S}_n^\ast.
\end{aligned}\right.
\end{eqnarray}
After fixing the initial conditions, we can solve equations (\ref{c3}) and (\ref{c4}). On the other hand, the adiabatic initial condition according to the inflationary theory may be written as\cite{6}
\begin{equation}
\epsilon\longrightarrow 0 \quad\mbox{or}\quad y\longrightarrow0 \quad :\quad \mathcal{R}_q\longrightarrow Nq^{-2+\frac{n_{s_0}}{2}}\quad\mbox{and}\quad\mathcal{S}_q\longrightarrow0.
\end{equation}
Where according to the observational data $ N\simeq10^{-5} $ and $ n_{s_0}\simeq0.96 $ \cite{20}. So under the
adiabatic initial condition, we have
\begin{eqnarray*}
&&\alpha=\beta=-1+\frac{n_{s_0}}{4},\\
&&\mathcal{S}_0\left(y \right)=\mathcal{S}_1\left(y \right)=0,\\
&&\mathcal{R}_0\left(y \right)=N\mathcal{H}_{eq}^{-2+\frac{n_{s_0}}{2}},\\
&&\mathcal{R}_1\left(y \right)=-\frac{16}{45}N\left[ \frac{1}{y}+\frac{13}{3\left( 3y+4\right) }-\frac{2\sqrt{y+1}\left( 3y+2\right)} {y\left( 3y+4\right) }+2\ln{ \frac{1+\sqrt{y+1}}{2} }-\frac{13}{12}\right]. 
\end{eqnarray*}
So, when the scales of the perturbations are not extremely smaller than the Hubble horizon, under the adiabatic initial condition the following approximation is appropriate
\begin{eqnarray*}
\left\{\begin{aligned}
&\mathcal{S}_0\left(y \right)\simeq0,\\ \\
&\mathcal{R}_q\left(y \right)\simeq Nq^{-2+\frac{n_{s_0}}{2}}-\frac{16N}{45\mathcal{H}_{eq}^2}\Bigg[ \frac{1}{y}+\frac{13}{3\left( 3y+4\right) }-\frac{2\sqrt{y+1}\left( 3y+2\right)} {y\left( 3y+4\right) }+2\ln{ \frac{1+\sqrt{y+1}}{2} }-\frac{13}{12}\Bigg]q^{\frac{n_{s_0}}{2}}.
\end{aligned}\right.
\end{eqnarray*}
On the other hand, the isocurvature initial condition may be written as
\begin{equation}
\epsilon\longrightarrow 0 \quad\mbox{or}\quad y\longrightarrow0 \quad :\quad \mathcal{S}_q\longrightarrow Mq^{-2+\frac{n_{iso_0}}{2}} \quad\mbox{and} \quad\mathcal{R}_q\longrightarrow\frac{y}{3y+4}\mathcal{S}_q . 
\end{equation}
where in accordance with the \textit{Liddle and Mazumdar model} \cite{24} $ n_{iso_0} \simeq 4.43$. It may be also considered $ M\simeq10^{-5} $ as similar as domain of the adiabatic perturbation. Unfortunately, in this case solving the recursive equations result in calculation of some enormous integrals, so we leave it. Nevertheless, it is possible to solve equations (\ref{b2}) and (\ref{b9}) numerically. We present the results in some figures. In Figures \ref{fig1} and \ref{fig2} the curve of $ \mathcal{R}_q $ for the adiabatic and isocurvature initial conditions have been plotted. 

%\begin{figure}[tbp]
%\centering % \begin{center}/\end{center} takes some additional vertical space
%\includegraphics[width=.45\textwidth,trim=0 380 0 200,clip]{img1.pdf}
%\hfill
%\includegraphics[width=.45\textwidth,origin=c,angle=180]{img2.pdf}
% "\includegraphics" is very powerful; the graphicx package is already loaded
%\caption{\label{fig:i} Always give a caption.}
%\end{figure}

%\begin{figure}[h]
%\begin{center}$
%\begin{array}{cc}
%\includegraphics[width=2.5in]{1.jpg} &
%\includegraphics[width=2.5in]{9.jpg}
%\end{array}$
%\end{center}
%\caption{my caption}
%\end{figure}

\begin{figure}[!htb]
\centering
\includegraphics[width=.6\textwidth]{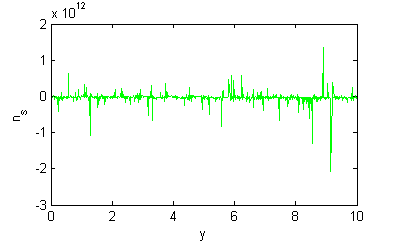}
\hfill
\includegraphics[width=.6\textwidth]{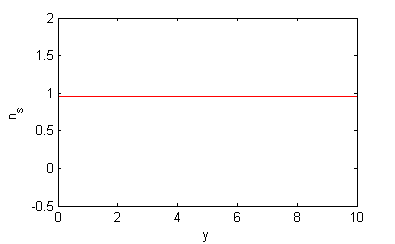}
\caption{Evolution of the adiabatic spectral index (spectral index of $ \mathcal{R}_q $) in a universe constructed from dust and radiation for the comoving wave number $ q=10^5 $ i.e. sub-horizon scales (up) and $ q=10^{-10} $ namely severe super-horizon modes (down) providing the adiabatic initial condition. We suppose $ n_{s_0}=0.96 $ and $ N $, the amplitude of $ \mathcal{R}_q $ at the end of inflation is roughly $ 10^{-5} $.}\label{fig3}
\end{figure}

\begin{figure}[!htb]
\centering
\includegraphics[width=.6\textwidth]{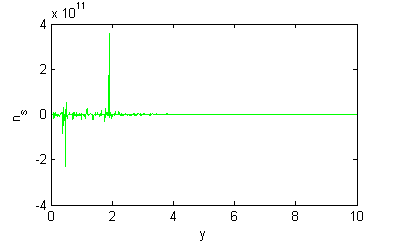}
\hfill
\includegraphics[width=.6\textwidth]{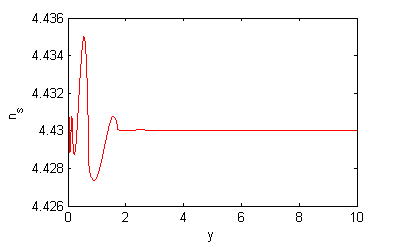}
\caption{The same as Figure \ref{fig3}, except the initial condition supposed to be isocurvature. We assumed the amplitude of $ \mathcal{S}_q $ at the end of inflation is about $ 10^{-5} $ and $ n_{iso_0}=4.43 $. Notice that $ n_s $ seems to be constant for $ y\gtrsim 4 $ even under isocurvature initial condition.}\label{fig4}
\end{figure}

\begin{figure}[!htb]
\centering
\includegraphics[width=.6\textwidth]{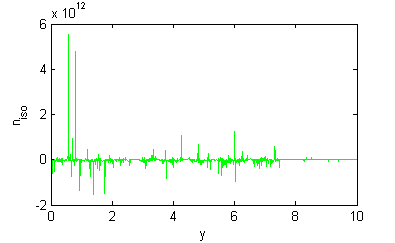}
\hfill
\includegraphics[width=.6\textwidth]{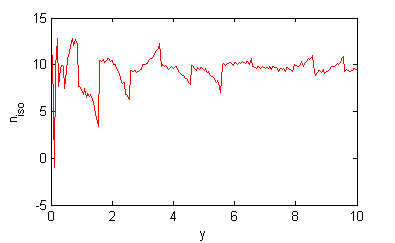}
\hfill
\includegraphics[width=.6\textwidth]{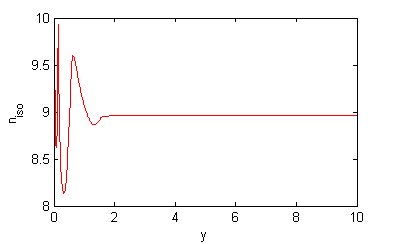}
\caption{Evolution of the isocurvature spectral index (spectral index of $ \mathcal{S}_q $) in a universe constructed from dust and radiation for the comoving wave number $ q=10^5 $ i.e. sub-horizon scales (up), $ q=10^{-5} $ (middle) and $ q=10^{-10} $ (down) namely super-horizon modes  providing the adiabatic initial condition. We suppose $ n_{s_0}=0.96 $ and $ N $, the amplitude of $ \mathcal{R}_q $ at the end of inflation is roughly $ 10^{-5} $. It is clear that $ n_{iso}$  for the adiabatic initial condition is roughly constant of course just for severe super-Hubble scales for which $ q\ll \mathcal{H} $ and $ y\gtrsim 2 $. }\label{fig5}
\end{figure}

\begin{figure}[!htb]
\centering
\includegraphics[width=.6\textwidth]{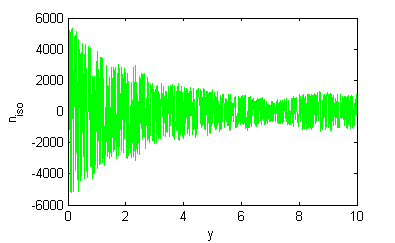}
\hfill
\includegraphics[width=.6\textwidth]{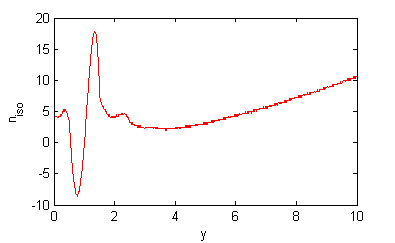}
\hfill
\includegraphics[width=.6\textwidth]{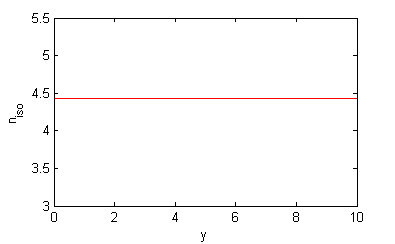}
\caption{The same as Figure \ref{fig5}, however for the isocurvature initial condition. We supposed the amplitude of $ \mathcal{S}_q $ at the end of inflation is about $ 10^{-5} $ and $ n_{iso_0}=4.43 $. It is clear that $ n_{iso} $ for the severe super-Hubble scales and isocurvature initial condition is constant.}\label{fig6}
\end{figure}

The spatial indices of the adiabatic and isocurvature perturbations respectively are defined as
\begin{eqnarray}
n_s\left( q\right)=4+\frac{q}{\mathcal{P}_\mathcal{R}\left( q\right)}\frac{\partial \mathcal{P}_\mathcal{R}\left( q\right)}{\partial q}=4+2\frac{q}{\mathcal{R}_q}\frac{\partial\mathcal{R}_q}{\partial q} ,\\
n_{iso}\left( q\right) =4+\frac{q}{\mathcal{P}_\mathcal{S}\left( q\right)}\frac{\partial \mathcal{P}_\mathcal{S}\left( q\right)}{\partial q}=4+2\frac{q}{\mathcal{S}_q}\frac{\partial\mathcal{S}_q}{\partial q},
\end{eqnarray}
where $ \mathcal{P}_\mathcal{R}\left( q\right) $ and $ \mathcal{P}_\mathcal{S}\left( q\right) $ are respectively the power spectrums of $ \mathcal{R}_q $ and $ \mathcal{S}_q $. The curves of $ n_s\left( q\right) $ and $ n_{iso}\left( q\right) $ in terms of $ y $ has plotted in Figures \ref{fig3}, \ref{fig4}, \ref{fig5} and \ref{fig6} for the adiabatic and isocurvature initial conditions respectively. It is clear that $ n_s $ and $ n_{iso} $ depend on $ q $ severely, so we may say the spectral indices are running. 

%%%%%%%%%%%%%%%%%%%%%%%%%%%%%%%%%%
\section{The Sachs-Wolfe effect: a new survey}
\label{sec:3}

Cosmic Microwave Background (CMB) was discovered in a study of noise backgrounds in a radio telescope by Penzias and Wilson in 1965 \cite{25}. Two years later, Sachs and Wolfe pointed out that the CMB must show the temperature anisotropy as a result of photon traveling in the perturbed universe \cite{26}. An important contribution in temperature anisotropy of CMB is called the ordinary Sachs-Wolfe effect which stems from the intrinsic temperature inhomogeneities on the last scattering surface and also the inhomogeneities of the metric at the time of last scattering. It can be shown that \cite{27,28,29}
\begin{equation}\label{d2}
\left[ \frac{\Delta T\left(\hat{\textbf{n}} \right) }{T_0}\right] _{\rm{S.W.}}=\zeta_R\left(t_L,\hat{\textbf{n}}r_L \right)+\Psi \left(t_L,\hat{\textbf{n}}r_L\right)+\Phi \left(t_L,\hat{\textbf{n}}r_L\right),
\end{equation}
where $ \zeta_R $ denotes the curvature perturbation of the radiation in the uniform density slices. Moreover, $ \hat{\textbf{n}} $ is the unit vector stands for the direction of observation. Notice all quantities are evaluated in last scattering surface. The corresponding scales are well outside the Hubble horizon has dominant imprint in the ordinary Sachs-Wolfe effect. In the multipole space, the ordinary Sachs-Wolfe effect is responsible for $ l\lesssim 30 $ \cite{5}. Moreover, its angular power spectrum has a plateau which is known as Sachs-Wolfe plateau. In the mixture of radiation and dust equation (\ref{d2}) reduces to
\begin{equation}\label{c5}
\left[ \frac{\Delta T\left(\hat{\textbf{n}} \right) }{T_0}\right] _{\rm{S.W.}}=\zeta_R\left(t_L,\hat{\textbf{n}}r_L\right)+2\Psi \left(t_L,\hat{\textbf{n}}r_L\right).
\end{equation}
Calculating of $ \zeta_R $ for a radiation-dust mixture is straightforward, because $ \mathcal{S}=3\left( \zeta_M-\zeta_R\right)  $. On the other hand, $ \zeta $ is the weighted average of $ \zeta_R $ and $ \zeta_R $ i.e.
\begin{equation}
\zeta=\frac{4}{3y+4}\zeta_R+\frac{3y}{3y+4}\zeta_M.
\end{equation}
Thus
\begin{equation}
\zeta_R=\zeta-\frac{y}{3y+4}\mathcal{S}.
\end{equation}
Moreover, we have
\begin{equation}
\zeta=\mathcal{R}+\frac{1}{3\left( \mathcal{H}^2-\mathcal{H}'\right) }\nabla^2\Psi.
\end{equation}
Consequently
\begin{equation}\label{c6}
\zeta_R=\mathcal{R}+\frac{4y^2}{3\mathcal{H}_{eq}^2\left(3y+4\right) }\nabla^2\Psi-\frac{y}{3y+4}\mathcal{S}.
\end{equation}
Returning equation (\ref{c6}) to equation (\ref{c5}) we find that
\begin{equation}\label{c7}
\left[ \frac{\Delta T\left(\hat{\textbf{n}} \right) }{T_0}\right] _{\rm{S.W.}}=\mathcal{R}\left(t_L,\hat{\textbf{n}}r_L\right)+2\Psi \left(t_L,\hat{\textbf{n}}r_L\right)-\frac{y_L}{3y_L+4}\mathcal{S} \left(t_L,\hat{\textbf{n}}r_L\right).
\end{equation}
Note that we omitted the term containing $ \nabla^2\Psi $ in equation (\ref{c7}), because in the ordinary Sachs-Wolfe the small scales has subdominant contribution. In the case of adiabatic initial condition equation (\ref{c7}) reduces to
\begin{equation}\label{c8}
\left[ \frac{\Delta T\left(\hat{\textbf{n}} \right) }{T_0}\right] _{\rm{S.W.}}^{\rm{Ad}}=\mathcal{R}\left(t_L,\hat{\textbf{n}}r_L\right)+2\Psi \left(t_L,\hat{\textbf{n}}r_L\right).
\end{equation}
On the other hand,
\begin{equation}
\frac{2}{3\left( \omega+1\right) \mathcal{H}}\Psi'+\frac{3\omega+5}{3\left( \omega+1\right) }\Psi=-\mathcal{R}.
\end{equation}
or
\begin{equation}\label{c9}
\Psi^\ast+\frac{5y+6}{2y\left( y+1\right)}\Psi=-\frac{3y+4}{2y\left(y+1 \right) }\mathcal{R}.
\end{equation}
Here $ \mathcal{R} $ is independent of $ y $, so equation (\ref{c9}) yields
\begin{equation}\label{d1}
\Psi=\frac{-9y^3-2y^2+8y+16-16\sqrt{y+1}}{15y^3}\mathcal{R}.
\end{equation}
Notice that
\begin{eqnarray*}
&\lim \Psi=-\frac{2}{3}\mathcal{R}\hspace{35pt} &\lim \Psi=-\frac{3}{5}\mathcal{R} 
\\&y\longrightarrow 0 & y\longrightarrow \infty
\end{eqnarray*}
which is expected for the pure radiation and pure dust respectively. By substituting equation (\ref{d1}) in equation (\ref{c8}) we find that
\begin{equation}
\left[ \frac{\Delta T\left(\hat{\textbf{n}} \right) }{T_0}\right] _{\rm{S.W.}}^{\rm{Ad}}=\left[2-\frac{15y_L^3}{9y_L^3+2y_L^2-8y_L-16+16\sqrt{y_L+1}} \right] \Psi \left(t_L,\hat{\textbf{n}}r_L\right).
\end{equation}
In addition, we have
\begin{equation}
y_L=\frac{1+z_{eq}}{1+z_L}=\frac{1+3263}{1+1091}=2.98.
\end{equation}
So
\begin{equation}
\left[ \frac{\Delta T\left(\hat{\textbf{n}} \right) }{T_0}\right] _{\rm{S.W.}}^{\rm{Ad}}= 0.4 \Psi \left(t_L,\hat{\textbf{n}}r_L\right).
\end{equation}
So the fudge factor of Sachs-Wolfe effect should be substitude by 0.4 which is greater than the ordinary 1/3 factor \cite{29,30,31}.\\
Finally, lets turn to the isocurvature initial condition. In this case we have $ \mathcal{R}=\frac{y}{3y+4}\mathcal{S} $, so
\begin{equation}
\left[ \frac{\Delta T\left(\hat{\textbf{n}} \right) }{T_0}\right] _{\rm{S.W.}}^{\rm{Iso}}=2 \Psi \left(t_L,\hat{\textbf{n}}r_L\right),
\end{equation}
which coincides with previous results completely \cite{29,30}.
%%%%%%%%%%%%%%%%%%%%%%%%%%%%%%%%%%
\section{Discussion}
\label{sec:4}

We have derived a neat equation for the evolution of comoving curvature perturbation and then have found its solutions for some simple cases. We showed that the Mukhanov-Sasaki equation is a special case of this equation. We also found its numerical solutions for the radiation and matter mixing by fixing two different adiabatic and isocurvature initial conditions. As we have seen, in this case the equation cannot be solved alone due to the presence of another unidentified quantity: entropy perturbation. So we coupled the equation to the Kodama-Sasaki equation. We also investigated the time evolution of the adiabatic and isocurvature spectral indices for both initial conditions separately. We showed that not only the curvature spectral index for adiabatic initial condition and severe super-Hubble scales is constant, but also the entropy spectral index for isocurvature initial condition and severe super-Hubble scales is constant too. It seems $ n_{iso} $ for $ y\gtrsim 2 $ and severe supe-Hubble scales is approximately flat regardless to the initial condition. Moreover, we found that $ n_s $ is roughly constant for $ y\gtrsim 4 $ regardless to the scale of perturbations even though the isocurvature initial condition has been chosen.  Finally, we re-investigated the ordinary Sachs-Wolfe effect and showed that the factor $ \frac{1}{3}  $ should be increased as amount $ \sim $0.06 by considering more actual situation. This is consistent by White and Hu pedagogical derivation of Sachs-Wolfe effect.

%\acknowledgments

%We are grateful to V. Barger, T. Han, and R. J. N. Phillips for
%doing the math in section~\ref{bozomath}.


\end{document}